\documentclass[aps,prd,amsfonts,amsmath,twocolumn]{revtex4}

\usepackage{color}
\usepackage{graphics,epsfig}

\usepackage{epstopdf}

\newcommand {\be} {\begin{equation}}
\newcommand {\ba} {\begin{eqnarray}}
\newcommand {\ee} {\end{equation}}
\newcommand {\ea} {\end{eqnarray}}

\begin{document}

\preprint{WM-07-105}

\title{Gravitational Form Factors of Vector Mesons in an AdS/QCD Model}

\author{Zainul Abidin and Carl E.\ Carlson}
\affiliation{
Department of Physics, College of William and Mary, Williamsburg, VA 23187, USA}

\date{\today}

\begin{abstract}
We calculate gravitational form factors of vector mesons using a holographic model of QCD. These provide restrictions on the generalized parton distributions of vector mesons, via the sum rules connecting stress tensor form factors to GPDs.  We concentrate on the traceless part of the stress tensor, which suffices to fix the momentum and angular momentum sum rules.  The vector mesons appear noticeably more compact measured by the gravitational form factors than by the charge form factor.
\end{abstract}

\maketitle

%
\section{Introduction}
%

In this paper we calculate gravitational form factors---form factors for the stress or energy-momentum tensor---of vector mesons using a hard-wall model of AdS/QCD.

There has been much interest in the AdS/CFT, or gauge/gravity, correspondence because it offers the possibility of relating nonperturbative quantities in theories akin to QCD in 4 dimensions to gravitationally coupled 5-dimensional theories that are treated perturbatively~\cite{Maldacena:1997re}.  Many applications have already been made; see references~\cite{Sakai:2004cn,Hirn:2005nr,Karch:2002sh,Erlich:2005qh,Da Rold:2005zs,deTeramond:2005su,Brodsky:2006uqa,Karch:2006pv,ArkaniHamed:2000ds,Grigoryan:2007vg,Grigoryan:2007my,Schafer:2007qy,Evans:2007sf,Erdmenger:2007cm} and other works cited in those references.   However, for objects of particular interest to hadron structure physics such as ordinary parton distribution functions, form factors, transverse momentum dependent parton distribution functions, and generalized parton distributions, there is a smaller body of work, particularly for the latter two topics.

Part of the interest in gravitational form factors comes because of their connections to generalized parton distributions (GPDs).  GPDs are an important metric of hadron structure, and can be loosely described as amplitudes for removing a parton from a hadron and replacing it with one of different momentum.  Moments of the GPDs are related to gravitational form factors.  In particular, one of the gravitational form factors measures the total angular momentum carried by partons, and historically the possibility to find the summed spin plus orbital angular momentum of particular constituents of hadrons is what keyed the current experimental and theoretical interest in GPDs~\cite{Ji:1996ek,Diehl:2003ny}.

The original AdS/CFT correspondence~\cite{Maldacena:1997re} related a strongly-coupled, large $N_c$, 4D conformal field theory  with  a weakly-coupled gravity theory on 5D AdS space. In QCD, $N_c$ is not large, nor is it a conformal field theory, as evidenced by the existence of hadrons with definite mass. Nonetheless, results obtained treating $N_c$ as large work surprisingly well, and one can argue that QCD behaves approximately conformally over wide regions of $Q^2$~\cite{Brodsky:2006uqa}.   The AdS/CFT correspondence has been studied in both a ``top-down'' approach, starting from string theory~\cite{Sakai:2004cn,Karch:2002sh}, and a ``bottom-up'' approach, which uses the properties of QCD to construct its 5D gravity dual theory~\cite{Erlich:2005qh,Da Rold:2005zs,deTeramond:2005su,Brodsky:2006uqa,Karch:2006pv}. We follow the latter approach, particularly as implemented in~\cite{Erlich:2005qh,Da Rold:2005zs}.  Some of the salient results include the meson spectrum, decay constants and electromagnetic form factors for both the rho and the pi~\cite{Grigoryan:2007vg,Grigoryan:2007wn,Kwee:2007dd,Grigoryan:2007my,Brodsky:2007hb}.

The model uses a sharp cut-off in the AdS space to simulate the breaking of conformal symmetry. The unperturbed metric and relevant slice of 5-dimensional AdS space is,
\be
ds^2=\frac{1}{z^2}(\eta_{\mu\nu}dx^\mu dx^\nu-dz^2),\qquad \varepsilon<z<z_0.
\ee
where $\eta_{\mu\nu}=\text{diag}(1,-1,-1,-1)$. The $z= \varepsilon$ wall, with $\varepsilon \to 0$ understood, corresponds to the UV limit of QCD, and the wall located at $z=z_0 \equiv 1/\Lambda_{\rm QCD}$ sets the scale for the breaking of conformal symmetry of QCD in the IR region.   [Lower case Greek indices will run from $0$ to $3$, and lower case Latin indices will run over $0,1,2,3,5$.]

Within AdS/CFT, every operator  $\mathcal{O}(x)$ in the 4D field theory corresponds to a 5D source field $\phi(x,z)$ in the bulk. Following the model proposed in~\cite{Erlich:2005qh,Da Rold:2005zs} two of the correspondences are
\ba
{J^a_L}^\mu(x) &\leftrightarrow& {A^a_L}^\mu(x,z) , \nonumber\\
{J^a_R}^\mu(x)&\leftrightarrow& {A^a_R}^\mu(x,z) ,
\ea
where ${J^a_L}^\mu=\bar{q}_L\gamma^\mu t^a q_L$ and ${J^a_R}^\mu=\bar{q}_R\gamma^\mu t^a q_R$ are the chiral flavor currents.

In Lagrangian formulations of general relativity, the source for the stress tensor $T_{\mu\nu}$ is the metric $g^{\mu\nu}$, whose variation is given in terms of $h^{\mu\nu}$.   We will use $h_{\mu\nu}$ in the Randall-Sundrum gauge, wherein $h_{\mu\nu}$ is transverse and traceless (TT) and also satisfies $h_{\mu z} = h_{zz} = 0$.  Variations of the metric tensor in a TT gauge will only give us the transverse-traceless part of the stress tensor.  This will, we shall see below, to uniquely determine 4 of the 6, for spin-1 particles, form factors of the stress tensor, including the two form factors that enter the momentum and angular momentum sum rules.  

The layout  of this paper is as follows.  Sections~\ref{sec:gravity},~\ref{sec:vector}, 
and~\ref{sec:ff}, obtain the form factors of the stress tensor using the AdS/CFT correspondence.  Specifically, Sec.~\ref{sec:gravity} obtains some necessary results for the purely gravitational parts of the theory, and Sec.~\ref{sec:vector} similarly focuses on the vector parts of the action, reviewing necessary results originally obtained in~\cite{Erlich:2005qh,Grigoryan:2007vg}.  Section~\ref{sec:ff} works out the three-point functions, and extracts from them the stress tensor matrix elements.

Section~\ref{sec:gpd} begins by giving the general expansion of the stress tensor for spin-1 particles, as constrained by the conservation law and symmetries, and then relates five combinations of stress tensor form factors to integrals over the five vector GPDs that exist for spin-1 particles~\cite{Berger:2001zb}.  One of these relations is the spin-1 version of the X. Ji or angular momentum sum rule~\cite{Ji:1996ek}.  The relations as a whole are a set of contraints upon the spin-1 GPDs.  We also determine the radius of the vector meson from the form factors that enter the momentum and angular momentum sum rules, and compare to the electromagnetic radius.
Some conclusions are offered in Sec.~\ref{sec:theend}.

%
\section{Gravity Sector}  \label{sec:gravity}
%

The action on the 5-dimensional AdS space is
\ba
S_{5D}&=&\int d^5 x \sqrt{g}\bigg\{ R+12			 \label{fullAction}	\\
&&+\text{Tr}\Big[  |DX|^2+3|X|^2-\frac{1}{4g_5^2}(F_L^2+F_R^2) \Big] \bigg\}   ,
					\nonumber
\ea
where $F_{mn}=\partial_m A_n-\partial_n A_m-i[A_m,A_n]$, $A_{L,R}=A^a_{L,R}t^a$, with Tr$(t^at^b)=\delta^{ab}/2$  and $D^m X=\partial^m X-i A_L^m X+iX A_R^m$.  We have only displayed fields needed for this paper, and we impose Dirichlet boundary conditions on the $z_0$ boundary.

In the purely gravitational part of the action,
\be
S_G=\int d^5 x \sqrt{g} (R+12),
\ee
the metric is perturbed from its AdS background according to
\be
ds^2=\frac{1}{z^2}((\eta_{\mu\nu}+h_{\mu\nu})dx^\mu dx^\nu-dz^2), \quad 0<z<z_0,
\label{eq:fullmetric}
\ee
where (so far) $h_{zz}=0$, $h_{z\mu}=0$ gauge choices have been used.
The linearized Einstein equations are
\ba
0&=&- {h}_{\mu\nu,zz}
	+\frac{3}{z} {h}_{\mu\nu,z}
		+ {h_{\mu\nu,\rho}}^\rho
			- 2 {h^\rho}_{(\mu,\nu)\rho} 
		\nonumber	\\
&&+\eta_{\mu\nu}( {\tilde{h}}{_{,zz}} - \frac{3}{z} \tilde{h}_{,z}
	- {\tilde{h}_{,\rho}}{} ^\rho
		+ {h_{\rho\sigma,}}^{\rho\sigma})+\tilde{h}_{,\mu\nu}
		\nonumber	\\
0&=&\tilde{h}_{,\mu z} - {{h_{\mu\nu,z}}^\nu}
		\nonumber	\\
0&=&\frac{3}{z}{\tilde{h}_{,z}}
	+ {\tilde{h}_{,\rho}}{}^\rho
		- {h_{\rho\sigma,}}^{\rho\sigma}
\ea
which come from the $\mu\nu$, $\mu z$, and  $zz$ sector of the Einstein equation. The trace of $h_{\mu\nu}$ is denoted by $\tilde{h}$. In transverse-traceless gauge, ${h}_{\mu\nu},^\nu=0$ and ${h}^\mu_\mu=0$. The equation of motion becomes
\be
-z^3\partial_z\big(\frac{1}{z^3}\partial_z {h}_{\mu\nu}\big)+\partial^\rho\partial_\rho {h}_{\mu\nu}=0.\label{ttgraviton}
\ee

The 4D Fourier transform of the solution is written as ${h}_{\mu\nu}(q,z)=h(q,z)h^0_{\mu\nu}(q)$. We require that $h(q,\epsilon)=1$ so that $h^0_{\mu\nu}(q)$ is the Fourier transform of the UV-boundary value of the graviton. The IR boundary condition becomes $\partial_z h(q,z_0)=0$. With this boundary condition, the surface term from the IR boundary obtained when varying the action vanishes.
One finds
\be
h(q,z)=\frac{\pi}{4}q^2z^2\bigg(\frac{Y_1(qz_0)}{J_1(qz_0)}J_2(qz)-Y_2(qz)\bigg).
\ee
For spacelike momentum transfer $q^2=-Q^2<0$ the solution is conveniently rewritten as
\be
\mathcal{H}(Q,z)=\frac{1}{2}Q^2z^2\bigg(\frac{K_1(Qz_0)}{I_1(Qz_0)}I_2(Qz)+K_2(Qz)\bigg).
\ee

Symmetry of the two-index tensor $T^{\mu\nu}$ implies that there are 10 independent components. Conservation of energy-momentum, $q_\mu T^{\mu\nu}=0$, reduces this to 6 independent components. $T^{\mu\nu}$ can be decomposed into transverse-traceless part $\hat{T}^{\mu\nu}$ with 5 independent component, which leaves the transverse-not-traceless part with one independent component given by $\tilde{T}^{\mu\nu}=\frac{1}{3}(\eta^{\mu\nu}-q^\mu q^\nu/q^2)T$, where $T$ is the trace of $T^{\mu\nu}$:
\be
T^{\mu\nu}=\hat{T}^{\mu\nu}+\frac{1}{3}(\eta^{\mu\nu}-\frac{q^\mu q^\nu}{q^2})T \,. 
	\label{EnMomDecomp}
\ee

\noindent This conserved operator couples only to a transverse source. A variation $h^0_{\mu\nu}$ which is transverse and traceless can couple only to $\hat{T}^{\mu\nu}$.

%
\section{Vector Sector}    \label{sec:vector}
%
Define the vector field $V=(A_L+A_R)/2$ and the axial-vector field $A=(A_L-A_R)/2$, and consider only the vector part of the action, Eq.~(\ref{fullAction}):
\be
S_V=\int d^5 x \sqrt{g}~ \text{Tr}\{-\frac{1}{2g_5^2}F_V^2\},    \label{vectorAction}
\ee
where $(F_V)_{mn}=\partial_m V_n-\partial_n V_m$ up to quadratic order in the action. The metric used in this section is non-dynamical, {\it i.e.,} only the unperturbed part of Eq.~(\ref{eq:fullmetric}).

In the $V_z=0$ gauge, the transverse part of the vector field satisfies the following equation of motion,
\be
\bigg(\partial_z(\frac{1}{z}\partial_z V_\mu^a(q,z))+\frac{q^2}{z}V_\mu^a\bigg)_\perp=0 . 
\label{vector}
\ee
The solution of this equation of motion can be written as
\be
{V_\perp}_\mu(q,z)=V(q,z)V^0_\mu(q) \,,
\ee
where $V^0_\mu(q)$ is the Fourier transform of the source of the 4D vector current operator ${J_V^a}_\mu=\bar{q}\gamma_\mu t^a q$. Current conservation, $q_\mu J^\mu_V=0$, requires that the source is transverse. Therefore only the transverse part of the UV-boundary of 5D vector field will be considered as the source of $J^\mu_V$.

$V(q,z)$ is called the bulk-to-boundary propagator for the vector field, and has boundary conditions $V(q,\epsilon)=1$ and $\partial_z V(q,z_0)=0$. The bulk-to-boundary propagator is
\be
V(q,z)=\frac{\pi}{2}zq\left(\frac{Y_0(qz_0)}{J_0(qz_0)}J_1(qz)-Y_1(qz)\right).
\ee
Evaluating the action, Eq.~(\ref{vectorAction}), on the solution leaves only the surface term
\be
S_V=\int \frac{d^4 q}{(2\pi)^4} {V^0}^\mu(q){V^0}_\mu(q)\left(-\frac{\partial_z V(q,z)}{2g_5^2z} \right)_{z=\epsilon}.      \label{VAction}
\ee

The Kaluza-Klein tower of the $\rho$ mesons can be obtained from the normalizable solutions of Eq.~(\ref{vector}) with $q^2=m_n^2$. The boundary conditions for $\psi_n(x,z)$, the n-th KK-mode $\rho$-meson's wave function, are $\psi_n(z=0)=0$ and $\partial_z \psi_n(z_0)=0$. The solutions are
\be
\psi_n=\frac{\sqrt{2}}{z_0 J_1(m_n z_0)}z J_1(m_n z)  \,,
\ee
and satisfy normalization conditions  $\int (dz/z)  \psi_n^2(z)=1$.

Using Green's function methods to solve Eq.~(\ref{vector}), one can show that the bulk-to-boundary propagator can be written in terms of a sum over the infinite tower of KK-modes of the $\rho$-meson as
\be
V(q,z)=-g_5\sum_n \frac{F_n\psi_n(z)}{q^2-m_n^2} , \label{BulkToBoundary}
\ee
where $F_n=(1/g_5)(-\frac{1}{z'}\partial_{z'}\psi_n(z'))|_{z'=\epsilon}$. Similar results can be obtained by incorporating the Kneser-Sommerfeld expansion of Bessel functions. The constant $F_n$ is the decay constant of the vector meson, defined by
\be
<0|J_\mu^a(0)|\rho_n^b(p)>=F_n\delta^{ab}\varepsilon_\mu(p). \label{DecayConstant}
\ee

This can be seen by calculating 2-point function of vector currents. Taking functional derivatives with respect to $V^0$ 
the 5D action in Eq.~(\ref{VAction}), and changing ${V^0}^\mu{V^0}_\mu$ to ${V^0}^\mu\Pi_{\mu\nu}{V^0}^\nu$ using the restriction that ${V^0}$ is transverse, one finds
\be
i \int d^4 x \, e^{iqx}\left<0 \right| {\mathcal T} J_\mu^a(x)J_\nu^b(0) \left| 0 \right>
	=\Sigma(q^2)\Pi_{\mu\nu}  \delta^{ab},\label{tpf}
\ee
where $\Pi_{\mu\nu}=\left(\eta_{\mu\nu}-q_\mu q_\nu/q^2\right)$ is the transverse projector, and
\be
\Sigma(q^2)=-\frac{\partial_z V(q,z)}{g_5^2z}\bigg|_{z=\epsilon}=\frac{1}{g_5^2}\sum_n\frac{({\psi'}_n(\epsilon)/\epsilon)^2}{q^2-m_n^2+i\varepsilon}.
\ee
Using Eq.~(\ref{DecayConstant}), the left hand side of Eq.~(\ref{tpf}) can be written as
\be
i \int d^4 x \, e^{iqx} \left< 0 \right| {\mathcal T} J_\mu^a(x)J_\nu^b(0) \left| 0 \right>
	=\sum_n \frac{F_n^2\delta^{ab}}{q^2-m_n^2+i\varepsilon}\Pi_{\mu\nu}	\,,
\ee
which confirms the the interpretation of $F_n$ in Eq.~(\ref{BulkToBoundary}) as the decay constant of the $\rho$-meson.

%
\section{Gravitational Form Factors of Vector Meson}    \label{sec:ff}
%

Stress tensor matrix elements of spin-1 particles defined by $\big<\rho_n^a(p_1)|T^{\mu\nu}(q)|\rho_n^b(p_2)\big>$ can be extracted from 3-point function
\be
\big< 0 \big|T \big( {J_a}^\alpha(x)  T^{\mu\nu}(y)  {J_b}^\beta(w)  \big)\big|0\big> .	\label{3pf}
\ee
The Fourier transform of the above 3-point function can be expressed as $\big<{J^a}^\alpha(-p_2) T^{\mu\nu}(q){J^b}^\beta(p_1)\big>$.  In order to pick up the correct term for the elastic stress tensor matrix elements, we apply the completeness relation
\be
\sum_{n}\frac{d^3p}{(2\pi)^3 2p^0 } \big|\rho^a_n(p)\big>\big<\rho^a_n(p)\big|=1
\ee
twice, then multiply by
\be
\varepsilon_\alpha^*(p_2,\lambda_2)\varepsilon_\beta(p_1,\lambda_1)\left(p_1^2-m_n^2\right)\left(p_2^2-m_n^2\right)\frac{1}{F_n^2} \,,
\ee
and take the limit $p_1^2\rightarrow m_n^2$ and $p_2^2\rightarrow m_n^2$.

Consider the following part of the full action, Eq.~(\ref{fullAction}),
\be
S_V= - \frac{1}{4g_5^2}\int d^5 x \sqrt{g} \,  g^{lm}g^{pn}F^a_{mn}F^a_{lp}.
\ee
Only $hVV$  terms contribute to the 3-point functions,
\be
\big< 0 \big| {\mathcal T} {J}^\alpha(x) \hat{T}^{\mu\nu}(y){J}^\beta(w)\big| 0 \big>	=
	\frac{  -  2 \, \delta^3 S}{\delta V^0_\alpha (x) \delta h^0_{\mu\nu}(y) \delta V^0_\beta(w)} ,
		\label{ttEnergyMom}
\ee
where the functional derivative is evaluated at $h^0=V^0=0$.

The relevant terms in the action that contribute to the 3-point function can be written as
\be
S_V \stackrel{\to}{=} \frac{1}{2g_5^2} 
	\int \frac{d^5 x}{z}\bigg(\eta^{\rho\gamma}\eta^{\sigma\delta} {h}_{\gamma\delta}
		\big[-F_{\sigma z}F_{\rho z}+\eta^{\alpha\beta}F_{\sigma\alpha}F_{\rho\beta}\big]
			\bigg)	\,,
\ee
The energy-momentum tensor from Eq.~(\ref{ttEnergyMom}) must be conserved and traceless. Therefore one may apply the transverse-traceless projector
\ba
\eta^{\rho\gamma}\eta^{\sigma\delta} {h}_{\gamma\delta}&&\rightarrow {h}_{\gamma\delta}\bigg[\left(\eta^{\rho\gamma}-\frac{q^\rho q^\gamma}{q^2}\right)\left(\eta^{\sigma\delta}-\frac{q^\sigma q^\delta}{q^2}\right)\nonumber\\
&&-\frac{1}{3}\left(\eta^{\rho\sigma}-\frac{q^\rho q^\sigma}{q^2}\right)\left(\eta^{\gamma\delta}-\frac{q^\gamma q^\delta}{q^2}\right)\bigg].
\ea

Taking the functional derivatives and then extracting the gravitational form factor from the 3-point function, one obtains
\ba
&&\left<\rho_n^a(p_2,\lambda_2)\big|\hat{T}^{\mu\nu}(q)\big|\rho_n^b(p_1,\lambda_1)\right>=
				\nonumber \\
&& \quad (2\pi)^4 \delta^{(4)}(q+p_1-p_2) \, \delta^{ab} \,
	\varepsilon^*_{2 \alpha} \varepsilon_{1 \beta}
				\nonumber\\[1ex]
&&\times\bigg[ - A(q^2)\bigg(4 q^{[\alpha} \eta^{\beta](\mu} p^{\nu)}
	+2\eta^{\alpha\beta}p^\mu p^\nu\bigg)				\nonumber\\
&&\quad-\ \frac{1}{2} \hat C(q^2)
\eta^{\alpha\beta}\bigg(q^2 \eta^{\mu\nu} - q^\mu q^\nu  \bigg)
				\nonumber\\
&&\quad +\  D(q^2) \bigg( q^2 \eta^{\alpha(\mu}\eta^{\nu)\beta}
	- 2q^{(\mu}\eta^{\nu)(\alpha}q^{\beta)}
		+ \eta^{\mu\nu} q^\alpha q^\beta \bigg)
				\nonumber\\
&&\quad-\ \hat F(q^2) \frac{q^\alpha q^\beta}{m_n^2}
	\bigg( q^2 \eta^{\mu\nu}- q^\mu q^\nu \bigg)\bigg] \,, 
				\label {ttEM}
\ea
where $p=(p_1+p_2)/2$, $q=p_2-p_1$, $a^{[\alpha} b^{\beta]} = (a^\alpha b^\beta - a^\beta b^\alpha)/2$, and 
$a^{(\alpha} b^{\beta)} = (a^\alpha b^\beta + a^\beta b^\alpha)/2$.
The invariant functions are given by
\ba
A(q^2)&=&Z_2   \,,	\nonumber \\
\hat C(q^2)&=&  \frac{1}{q^2} 
	\bigg(\frac{4}{3}Z_1+\big( q^2 - \frac{8m_n^2}{3}\big)Z_2\bigg), \nonumber \\
D(q^2)&=& \frac{2}{q^2} Z_1+\left( 1- \frac{2m_n^2}{q^2} \right)Z_2, \nonumber \\
\hat F(q^2)&=& \frac{4m_n^2}{3q^4}  \left( Z_1 - m_n^2Z_2 \right),
\ea
with
\ba
Z_1&=&\int \frac{dz}{z} \mathcal{H}(Q,z)\partial_z\psi_n \partial_z\psi_n \,,
			\nonumber \\
Z_2&=&\int \frac{dz}{z} \mathcal{H}(Q,z) \psi_n \psi_n \,,
\ea
for spacelike momentum transfer.

The matrix element of $\hat T^{\mu\nu}$ in Eq.~(\ref{ttEM}) is indeed traceless.  It is not traceless term by term, but rather is written in a form that allows easy contact with the general expression for spin-1 matrix elements of the stress tensor, to be given shortly.

The difference between $\hat T^{\mu\nu}$, the traceless part of the stress tensor, and the full stress tensor is a term proportional to $(\eta^{\mu\nu} - q^\mu q^\nu / q^2)$, shown in Eq.~(\ref{EnMomDecomp}).  Adding such a term can only affect the terms $\hat C(q^2)$ and $\hat F(q^2)$ in Eq.~(\ref{ttEM}).  The form factors $A(q^2)$ and $D(q^2)$ will not change.

%
\section{Sum rules for the GPDs}     \label{sec:gpd}
%

%
\subsection{Stress tensor}     
%

The stress tensor is a symmetric and transverse two-index object, and is even under parity and time reversal.   As such has six independent components, which can be composed as a spin-2 operator plus a spin-0 operator.  Its matrix elements with spin-1 particles may in general be expanded in terms of six Lorentz structures multiplying six scalar functions,
\begin{align}
& \left\langle p_2,\lambda_2 \right|  T^{\mu\nu} \left| p_1, \lambda_1 \right\rangle = \varepsilon^*_{2\alpha} \varepsilon_{1\beta} 
						\nonumber \\
&	\times	\Big\{ -2 A(q^2) \eta^{\alpha\beta} p^\mu p^\nu
	- 4 \big(  A(q^2)+B(q^2)  \big)
	q^{[\alpha} \eta^{\beta](\mu} p^{\nu)}
						\nonumber \\
&	+ \frac{1}{2} C(q^2) \eta^{\alpha\beta} \left( q^\mu q^\nu - q^2 \eta^{\mu\nu} \right)
						\nonumber \\
&	+ D(q^2) \left[ q^\alpha q^\beta \eta^{\mu\nu} -2 q^{(\alpha} \eta^{\beta)(\mu} q^{\nu)}
					+ q^2 \eta^{\alpha(\mu} \eta^{\nu)\beta} \right]
						\nonumber \\
&	+ E(q^2) \frac{q^\alpha q^\beta}{m_n^2} p^\mu p^\nu
	+ F(q^2) \frac{q^\alpha q^\beta}{m_n^2} \left( q^\mu q^\nu - q^2 \eta^{\mu\nu} \right) \Big\}\,.
\end{align}

The second listed component has coefficient $(A+B)$ to notationally match the corresponding expansion for spin-1/2 particles.  Four of the scalar functions were given from the gauge-gravity correspondence in the last section, and one also learns
\be
B(q^2) = E(q^2) = 0. 
\ee 

%
\subsection{Vector GPDs for spin-1 particles}     
%

For a spin-1 particle, there are five vector GPDs, defined by~\cite{Berger:2001zb}
\ba
&& \int \frac{p^+ dy^-}{2\pi} e^{ixp^+y^-} \times 
						\nonumber \\
&& \quad	\left\langle p_2,\lambda_2 \right|  \bar q(-\frac{y}{2}) \gamma^+  
		q(\frac{y}{2})\left| p_1, \lambda_1 \right\rangle_{y^+=0, y_\perp = 0}
						\nonumber \\
&& = - 2 (\varepsilon_2^* \cdot \varepsilon_1) p^+ H_1
	- \left( \varepsilon_1^+ \, \varepsilon_2^* \cdot q 
		- {\varepsilon_2^+}^* \, \varepsilon_1 \cdot q \right) H_2
						\nonumber \\
&& \quad + \ q\cdot \varepsilon_1 \, q\cdot \varepsilon_2^* \frac{p^+}{m_n^2} H_3
	- \left( \varepsilon_1^+ \, \varepsilon_2^* \cdot q 
		+ {\varepsilon_2^+}^* \, \varepsilon_1 \cdot q \right) H_4
						\nonumber \\
&& \quad +\ \left( \frac{m_n^2}{(p^+)^2} \varepsilon_1^+ \, {\varepsilon_2^+}^*
		+ \frac{1}{3} (\varepsilon_2^* \cdot \varepsilon_1) \right) 2 p^+ \, H_5 \,.
\ea

\noindent Each of the GPDs has arguments, $H_i = H_i(x,\xi,t)$, where $q^+ = -2\xi p^+$ and $t = q^2$, and they are related to form factors for spin-1 particles by
\begin{align}
\int_{-1}^1 dx \, H_i(x,\xi,t) &= G_i(t) ,	 	&i&=1,2,3,  \nonumber
						\\
\int_{-1}^1 dx \, H_i(x,\xi,t) &= 0 ,		&i&=4,5	\,.
\end{align}

\noindent The form factors are defined by the matrix elements of the vector current,
\ba
\left\langle p_2,\lambda_2 \right|  \bar q(0) \gamma^\mu
	q(0)\left| p_1, \lambda_1 \right\rangle 
		&=& -  (\varepsilon_2^* \cdot \varepsilon_1) 2p^\mu G_1(t)
			\nonumber \\[1ex]
	&& \hskip -4em - \left( \varepsilon_1^\mu \, \varepsilon_2^* \cdot q 
		- {\varepsilon_2^\mu}^* \, \varepsilon_1 \cdot q \right) G_2(t)
						\nonumber \\
&& \hskip -4em +\, q \cdot \varepsilon_1 \, q\cdot \varepsilon_2^* \frac{2p^\mu}{m_n^2} G_3(t) \,.
\ea
The $G_i$ are in turn related to the charge, magnetic, and quad\-rupole form factors by ($\eta = -t/(4m_n^2))$~\cite{Arnold:1979cg},
\ba
G_1 &=& G_C - \frac{2}{3} \eta G_Q	,		\nonumber \\
G_2 &=& G_M	,						\nonumber \\
(1+ \eta) G_3 &=& - G_C +G_M + \left(1+\frac{2}{3} \eta\right) G_Q  ,
\ea
normalized by $G_C(0)=1$, $G_M(0) = \mu_d$ (magnetic moment in units $(2m_n)^{-1}$), and $G_Q(0) = Q_d$ (quadrupole moment in units $m_n^{-2}$).

%
\subsection{Sum rules}     
%

For spin-1/2 constituents, 
\be
T^{++}(y) = \frac{i}{2} \bar q(y) \gamma^+ \overleftrightarrow \partial^+ q(x)
\ee
or,
\begin{align}
T^{++}(0) &= (p^+)^2 \int x dx  \nonumber \\
	&\times
	\int \frac{dy^-}{2\pi} e^{ixp^+y^-}
	\left[  \bar q(-\frac{y}{2}) \gamma^+  q(\frac{y}{2})\right]_{y^+=0, y_\perp = 0} \,,
\end{align}
so that there is a direct relation between the scalar functions in the stress tensor matrix elements and integrals over the GPDs.  These read,
\begin{align}
\int_{-1}^1 x dx \, H_1(x,\xi,t) &= A(t) - \xi^2 C(t) + \frac{t}{6m_n^2} D(t)	,	\nonumber \\
\int_{-1}^1 x dx \, H_2	(x,\xi,t) &= 2 \left( A(t)+B(t) \right)			,	\nonumber \\
\int_{-1}^1 x dx \, H_3(x,\xi,t) &= E(t) + 4 \xi^2 F(t)				,	\nonumber \\
\int_{-1}^1 x dx \, H_4(x,\xi,t) &= -2 \xi D(t)					,	\nonumber \\
\int_{-1}^1 x dx \, H_5(x,\xi,t) &= + \frac{t}{2m_n^2} D(t)				.	
\end{align}

\noindent (By time reversal invariance, $H_4$ is odd in $\xi$ while the other vector GPDs are even in $\xi$~\cite{Berger:2001zb,Diehl:2003ny}.)

\begin{figure}[ht]
\vglue -10 mm
\includegraphics[width = 3.35 in]{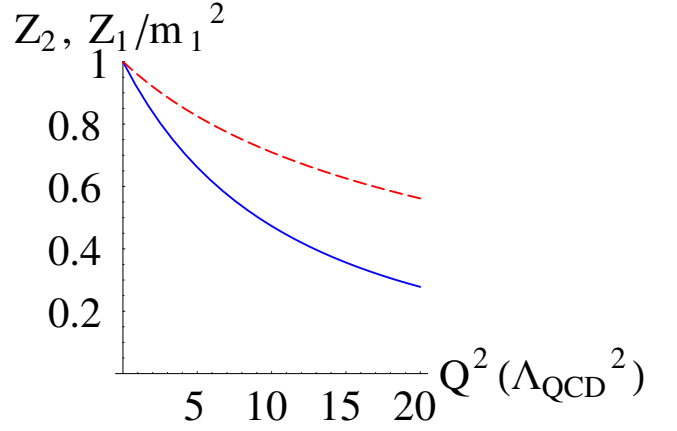}
\vglue -10 mm
\caption{Plot of $Z_2$ (solid blue curve) and $Z_1/M^2$ (dashed red curve), with momentum transfer in units of $\Lambda_\text{QCD} = 1/z_0$.}
\label{fig:Zplots}
\end{figure}

The relations between the stress tensor and the momentum and angular mometum operators lead to the sum rules (for any spin)~\cite{Ji:1996ek},
\ba
2 p^0 p^\mu \delta_{\lambda_1\lambda_2} &=& 
	\left\langle p,\lambda_2 \right|  T^{0\mu} \left| p, \lambda_1 \right\rangle
					\nonumber \\
2 p^0 \lambda_1 \delta_{\lambda_1\lambda_2} &=&
	i \bigg\{ \frac{\partial}{\partial q^x} 
		\left\langle p_2,\lambda_2 \right|  T^{02} \left| p_1, \lambda_1 \right\rangle
					\nonumber \\
&&	- 	\frac{\partial}{\partial q^y} 
		\left\langle p_2,\lambda_2 \right|  T^{01} \left| p_1, \lambda_1 \right\rangle
	\bigg\}_{q=0} \,,
\ea
where the latter is written for $\vec p$ in the $z$-direction.  Applied to spin-1 particles, this gives the normalizations 
\ba
A(0) &=& 1  \,,		\nonumber \\
A(0) + B(0) &=& \left( J_z \right)_{\rm max} = 1 \,.
\ea

When connected to the GPDs, the first of these is just the momentum sum rule,
\be
\int_{-1}^1 x dx \, H_1(x,0,0) = 1	,
\ee
and the second gives the spin-1 version of the X. Ji sum rule~\cite{Ji:1996ek}
\be
\int_{-1}^1 x dx \, H_2	(x,0,0) = 2 (J_z)_\text{max} = 2	.					
\ee
(Recall that $0 \le \xi \le \sqrt{-t/(4M^2-t)}$.)

Away from $t=0$, the gauge-gravity correspondence has led to a set of constraints, of which we will explicitly quote,
\begin{align}
&\int_{-1}^1 x dx \, \left(H_1(x,0,t) - \frac{1}{3}H_5(x,0,t) \right) = Z_2(t)	,	\nonumber \\
&\int_{-1}^1 x dx \, H_2	(x,0,t) = 2 Z_2(t)	,
\end{align}
where the $Z_i(t)$ are explicitly known, and are shown graphically in Fig.~\ref{fig:Zplots}.

%
\subsection{Radii}     
%

The RMS radius obtained from the gravitational form factor $A(q^2)$ is defined from
\be
\left\langle r^2 \right\rangle_{\rm grav} = -6 \left. \frac{\partial A}{\partial Q^2} \right|_{Q^2=0} .
\ee
For small $Q^2$ we expand
\be
{\cal H}(Q,z) = 1 - \frac{ Q^2 z^2 }{4} \left( 1 - \frac{z^2}{2z_0^2} \right) + {\cal O}(Q^4z^4) ,
\ee
and for the lightest vector meson obtain
\be
\left\langle r^2 \right\rangle_{\rm grav} = \frac{3.24}{m_1^2} = 0.21 {\rm\ fm}^2 ,
\ee
where we identified $m_1 = m_\rho$.

This is quite small.  The charge radius of the rho-meson obtained from AdS/CFT in~\cite{Grigoryan:2007vg} (and verified by us) is $\left\langle r^2 \right\rangle_C = 0.53 {\rm\ fm}^2$.  Similar charge radius results are obtained by a Dyson-Schwinger equation study~\cite{Bhagwat:2006pu} and from lattice gauge theory~\cite{Hedditch:2007ex}.  The result indicates that while the charge is spread over a certain volume, the energy that contributes to the mass of the particle is concentrated in a smaller kernel.

%
\section{Conclusions}     \label{sec:theend}
%

We have worked out the gravitational form factors of the vector mesons using the AdS/CFT correspondence, and have given the sum rules connecting the gravitational form factors, which can also be called stress tensor or energy-momentum tensor form factors, to the vector meson GPDs.

A striking numerical result is the smallness of the vector meson radius as obtained from $A(q^2)$, the gravitational form factor that enters the momentum sum rule.  This suggests that the energy that makes up the mass of the meson is well concentrated, with the charge measured by the electromagnetic form factors spreading more broadly.

Extensions of the present work include considering AdS/CFT implications for stress tensor form factors of the pion, whose electromagnetic properties were studied the AdS/CFT context in~\cite{Kwee:2007dd,Grigoryan:2007wn,Brodsky:2007hb};  flavor decompositions~\cite{Sakai:2004cn,Karch:2002sh,Kruczenski:2003be} of the stress tensor, particularly in connection with the angular momentum sum rule; and applying the present considerations to nucleons~\cite{deTeramond:2005su}.  We hope to return to these topics, but for the moment they lie beyond the scope of this paper.


\begin{acknowledgments}
We thank Josh Erlich and Marc Vanderhaeghen for conversations and suggestions, and the National Science Foundation for support under grant PHY-0555600.
\end{acknowledgments}

\end{document}